\documentclass[conference]{IEEEtran}
\IEEEoverridecommandlockouts
\usepackage{cite}
\usepackage{amsmath,amssymb,amsfonts}
\usepackage{algorithmic}
\usepackage[colorlinks=true,linkcolor=black,citecolor=black,urlcolor=black]{hyperref}

\usepackage{graphicx}
\usepackage{textcomp}
\usepackage{xcolor}
\usepackage{CJKutf8}

\def\BibTeX{{\rm B\kern-.05em{\sc i\kern-.025em b}\kern-.08em
		T\kern-.1667em\lower.7ex\hbox{E}\kern-.125emX}}

\begin{document}

\begin{CJK*}{UTF8}{gbsn}
	\title{An Enhanced Encoder-Decoder Network Architecture for Reducing Information Loss in Image Semantic Segmentation 
	}

     \author{
            \IEEEauthorblockN{1\textsuperscript{st} Zijun Gao}
            \IEEEauthorblockA{
                \textit{Northeastern University} \\
                Boston, USA \\
                zjg.elaine@gmail.com}
            \and
             \IEEEauthorblockN{2\textsuperscript{nd} Qi Wang}
            \IEEEauthorblockA{
                \textit{Northeastern University}\\
                Boston, USA \\
                bjwq2019@gmail.com}
           
            \and
            \IEEEauthorblockN{3\textsuperscript{rd} Taiyuan Mei}
           \IEEEauthorblockA{
                \textit{Northeastern University} \\
                Boston, USA \\
               taiyuanmei0824@gmail.com}
                
            \and
            \IEEEauthorblockN{4\textsuperscript{th} Xiaohan Cheng}
         \IEEEauthorblockA{
                \textit{Northeastern University } \\
                Boston, USA  \\
                Cheng.xiaoh@northeastern.edu}
            \and
            \IEEEauthorblockN{5\textsuperscript{th} Yun Zi}
            \IEEEauthorblockA{
                \textit{Georgia Institute of Technology} \\
                Chicago, USA   \\
                yzi9@gatech.edu}
     
 \and
            \IEEEauthorblockN{6\textsuperscript{th} Haowei Yang}
            \IEEEauthorblockA{
                \textit{University of Houston} \\
                Houston, USA   \\
                yanghaowei09@gmail.com}

           }
	
	\maketitle
	
	\begin{abstract}
The traditional SegNet architecture commonly encounters significant information loss during the sampling process, which detrimentally affects its accuracy in image semantic segmentation tasks. To counter this challenge, we introduce an innovative encoder-decoder network structure enhanced with residual connections. Our approach employs a multi-residual connection strategy designed to preserve the intricate details across various image scales more effectively, thus minimizing the information loss inherent to down-sampling procedures. Additionally, to enhance the convergence rate of network training and mitigate sample imbalance issues, we have devised a modified cross-entropy loss function incorporating a balancing factor. This modification optimizes the distribution between positive and negative samples, thus improving the efficiency of model training. Experimental evaluations of our model demonstrate a substantial reduction in information loss and improved accuracy in semantic segmentation. Notably, our proposed network architecture demonstrates a substantial improvement in the finely annotated mean Intersection over Union (mIoU) on the dataset compared to the conventional SegNet. The proposed network structure not only reduces operational costs by decreasing manual inspection needs but also scales up the deployment of AI-driven image analysis across different sectors.
\end{abstract}
	
\section{Introduction}
Semantic segmentation is a crucial task in the field of computer vision\cite{lu2023scaling}, with applications spanning from E-commerce webpage Recommendation\cite{Zhao_Liu_Xu_Xiao_Li_2024} to medical image analysis\cite{xiao2024convolutional}\cite{zhang2024optimization}. The goal of semantic segmentation is to partition an image into segments that correspond to different object classes. Traditional methods, including the widely-used SegNet architecture\cite{badrinarayanan2017segnet}, have laid a robust foundation for addressing these challenges. However, these approaches often suffer from significant information loss during the down-sampling process in the encoder phase, negatively impacting the accuracy of the segmentation.

Recognizing the limitations inherent in existing architectures, this paper introduces an innovative encoder-decoder network structure enhanced with residual connections designed to reduce information loss and improve segmentation accuracy. The core idea revolves around the integration of a multi-residual connection strategy that aids in preserving detailed information across various image scales, which is typically lost in traditional methods. By maintaining these details, our approach enhances the network's ability to perform accurate segmentation, even in complex visual scenes.

Furthermore, the challenges of network convergence and sample imbalance, which frequently plague the training of deep learning models, are addressed through the introduction of a modified cross-entropy loss function. This novel loss function incorporates a balancing factor that optimizes the distribution between the positive and negative samples, significantly refining the training efficiency and stability of the model.

In this paper, we will detail the development of our enhanced encoder-decoder network architecture, including the rationale behind the design choices and the specific implementations of the multi-residual connections and modified loss function. Experimental results demonstrate that our proposed model not only significantly reduces information loss but also achieves superior segmentation accuracy compared to the conventional SegNet architecture. This improvement is quantified through metrics such as the mean Intersection over Union (mIoU), highlighting our model's effectiveness in a practical, real-world context.

By reducing the need for manual inspection and facilitating the deployment of AI-driven image analysis, the proposed network structure offers promising prospects for enhancing operational efficiencies across various sectors. Through rigorous testing and validation on multiple datasets, we establish the robustness and scalability of our approach, setting a new standard for image semantic segmentation tasks.

	\section{ALGORITHM AND MODEL}
  
        \subsection{ SegNet Model}\label{sec:objectives-challenges}

SegNet is  a deep learning model primarily designed for image segmentation tasks, which is crucial in medical imaging\cite{wang2024investigating}\cite{yan2024survival}\cite{zhao2017lung} for tasks such as identifying and delineating regions\cite{10405429}\cite{liu2021comparison}, like tumors or various anatomical structures in MRI scans\cite{gong2024research}, CT scans, and other medical images.
The SegNet model consists of an encoder network and a corresponding decoder network, as shown in Figure 1.

  \begin{figure}[htbp]
        \centering
        \includegraphics[width=1\linewidth]{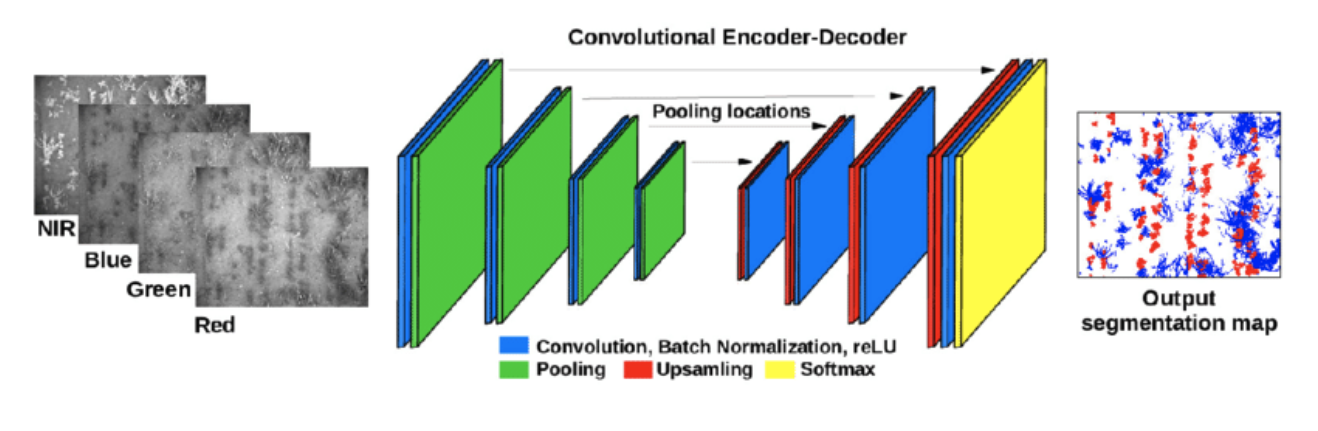}
        \caption{SegNet Model Architecture}
        \label{fig:SegNet Model Architecture}
        \end{figure}

The encoder network comprises convolutional layers, batch normalization layers,  and pooling layers. These layers extract features using same-padding convolution, normalize the data, and accelerate convergence with ReLU activation\cite{daubechies2022nonlinear}. Max-pooling layers record the positions of maximum values, providing robustness through translation invariance, though they reduce feature map size and spatial information. SegNet addresses this by storing only the max-pooling indices.

The decoder maps the encoded object and position information to specific pixels, up-samples the reduced feature maps, and refines object shapes using convolution, compensating for detail loss from the encoder’s pooling layers\cite{yao2023ndc}. Up-sampling layers double the feature map size based on the max-pooling indices, with other positions set to zero. The final output is fed into a softmax classifier for pixel-level classification\cite{wang2023self}.

SegNet’s innovation is using max-pooling indices from the encoder for unpooling in the decoder, reducing model parameters compared to FCN’s bilinear interpolation. This approach allows SegNet to achieve robust segmentation performance with balanced memory usage and accuracy, enhancing boundary delineation. SegNet also demonstrates superior inference time and memory efficiency compared to other architectures.

        \subsection{Model Establishment} 

This study designs a network structure based on the SegNet encoder-decoder architecture, incorporating residual connections for enhanced semantic segmentation\cite{li2024research}. For standard photographs, shallow CNNs capture more boundary and texture information\cite{xuxu2024research}, while deep CNNs extract higher-level abstract features. Combining both shallow and deep features is essential for improving semantic segmentation accuracy. While deepening and widening the network can enhance segmentation precision, it also introduces parameter burden and redundancy. Hence, residual connections and concatenation operations are used to effectively integrate shallow visual features with deep semantic features, with minimal additional parameters.

As depicted in Figure 2, the input image of size H×W produces an output of the same dimensions. Blue boxes represent convolutional layers followed by linear activation functions and batch normalization. The numbers within the boxes indicate the size and number of feature maps post-operation. During the encoder stage, the image is down-sampled thrice, reducing it to 1/8 of its original size and generating 256 feature maps. During the decoder stage, the feature maps are up-sampled to H×W, and the semantic category probabilities of each pixel are output via the softmax function.
  \begin{figure}[htbp]
        \centering
        \includegraphics[width=1\linewidth]{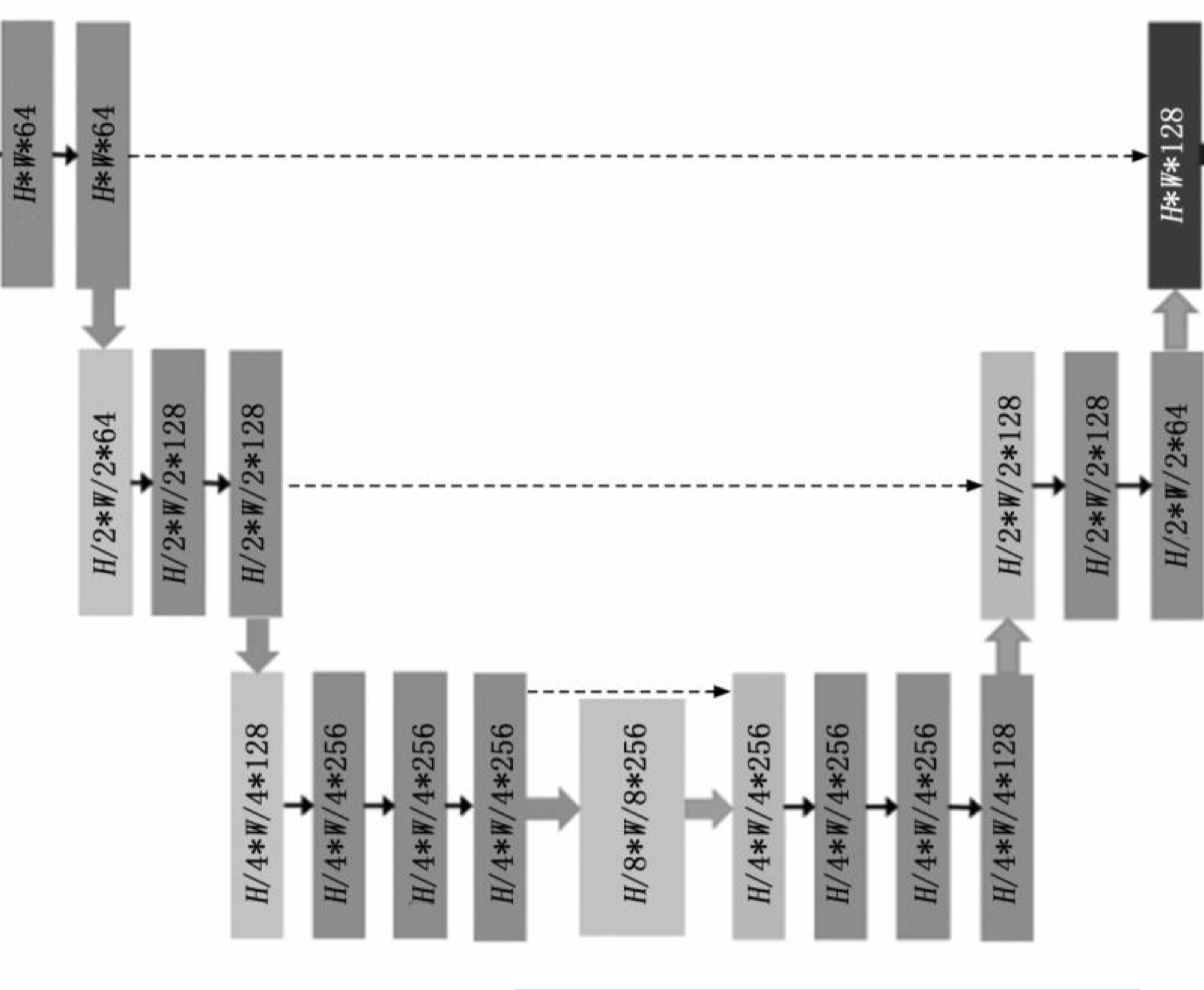}
        \caption{SegNet Model Architecture}
        \label{fig:Residual Connections for Enhanced Semantic Segmentation}
        \end{figure}
In the SegNet recovery stage, the down-sampled feature maps contain abundant feature information. However, due to network constraints, up-sampling sparse feature maps cannot generate dense feature maps, resulting in the loss of crucial information and suboptimal segmentation accuracy. Our network combines max-pooling indices and residual connections, feeding shallow feature maps from the encoder into the decoder's nonlinear up-sampling stage. Deconvolution then produces dense feature maps, preserving the original image's color, texture, and boundaries.

An input image undergoes the following steps in the improved SegNet network:

\begin{itemize}
    \item Convolution produces feature maps of H×W×64, denoted as F1.
    \item Down-sampling to H/2×W/2×64, followed by convolution to H/2×W/2×128, denoted as F2.
    \item Down-sampling to H/4×W/4×128, followed by convolution to H/4×W/4×256, denoted as F3.
    \item Down-sampling to H/8×W/8×256, denoted as F4.
    \item Up-sampling to H/4×W/4×256, denoted as F'3, calculated as F'3 = Fuse(PI(F1), F3). Deconvolution to H/4×W/4×128, denoted as De(F3).
    \item Up-sampling to H/2×W/2×128, denoted as F'2, calculated as F'2 = Fuse(PI(F2), De(F'3)). Deconvolution to H/2×W/2×64, denoted as De(F2).
    \item Up-sampling De(F2) to H×W resolution, combined with F1 through concatenation to H×W×128, denoted as F4, calculated as F4 = Conc(Fuse(PI(F1), De(F2)), F1).
    \item The softmax function assigns each pixel a category, outputting the semantic segmentation result.
\end{itemize}

        \subsection{Model Training}

The training process for the designed network model is as follows:

Preprocess the dataset and split it into training and validation sets.
Input the preprocessed data into the initialized semantic segmentation network model.
Iteratively update model parameters to minimize cross-entropy loss until convergence and minimal loss are achieved.
Output the optimal network model and parameters.
Figure 3 illustrates the network training flowchart.

  \begin{figure}
        \centering
        \includegraphics[width=1\linewidth]{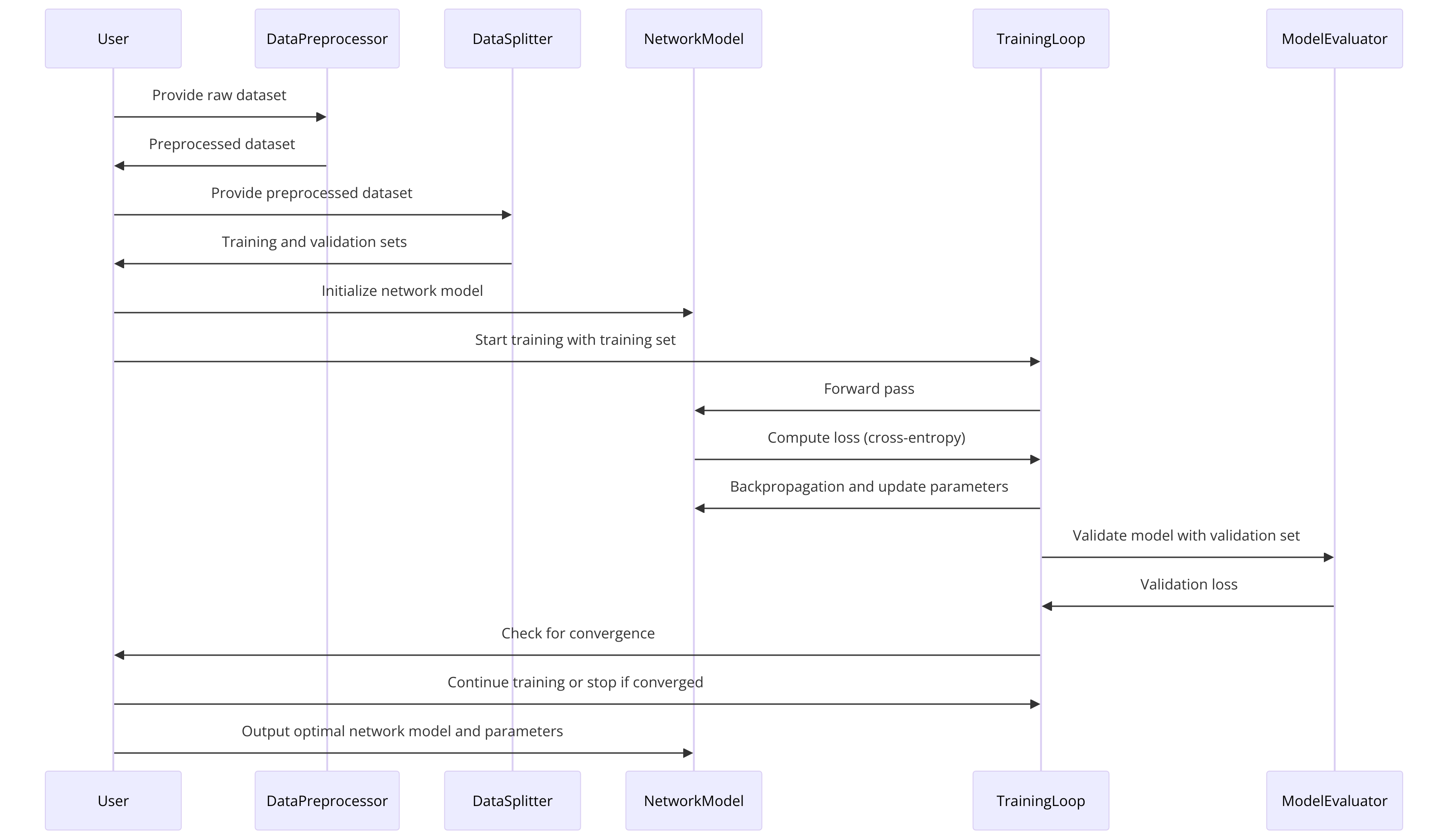}
        \caption{Training process}
        \label{fig:Training process}
        \end{figure}

    \subsection{Improved Cross-Entropy Loss Function} 
The standard cross-entropy loss function assigns equal weight to all samples. In cases of imbalanced positive and negative samples, the dominance of numerous easy negative samples can overshadow the impact of a few hard and positive samples, leading to reduced accuracy. To address this issue, we introduce a balancing factor within the range [0, 1].
\begin{equation}
    \beta =
\begin{cases} 
\beta & \text{if } (y = 1) \\
1 - \beta & \text{otherwise}
\end{cases}
\end{equation}

The improved cross-entropy loss function (B-CE) is formulated as follows:
\begin{equation}
    CE(p, y) = -\beta \log(p_t)
\end{equation}

By incorporating the balancing factor, the cross-entropy loss function achieves faster convergence than the standard version. This enhancement is particularly effective in optimizing class-imbalanced pixel distributions, thereby improving overall convergence efficiency.

        \section{Experimental Results and Analysis}
        \subsection{Experimental Environment}

The experiments were conducted on an advanced system running Windows 10 Professional, equipped with 32GB RAM, an Intel(R) Core i9-10980HK processor at 2.40GHz, and a NVIDIA  2080 8GB GPU. Matlab 2020b was used as the experimental platform, utilizing MatconvNet and Visual C++ 2017 for constructing the deep learning network model. Model training and testing were carried out in a CUDA 10.2-based GPU environment, ensuring enhanced processing speeds and computational efficiency.
        \subsection{Evaluation Metrics}

Intersection over Union (IoU) \cite{rezatofighi2019generalized} is a widely used metric for assessing the performance of object detection and semantic segmentation algorithms. It is computed as the ratio of the area of overlap between the predicted bounding box and the ground truth bounding box to the area of their union. IoU is straightforward and applicable to any task with a predicted range output. Figure 4 illustrates the mathematical concept of IoU.

Ideally, the candidate bounding box set completely overlaps with the ground truth, resulting in an IoU of 1, which indicates perfect prediction accuracy. The IoU formula is:

\begin{equation}
IoU = \frac{area(C) \cap area(G)}{area(C) \cup area(G)}
\end{equation}

Typically, an IoU threshold of 0.5 is used to determine the accuracy of the predicted bounding box. Higher IoU values indicate more precise bounding boxes.
\begin{figure}
    \centering
    \includegraphics[width=0.5\linewidth]{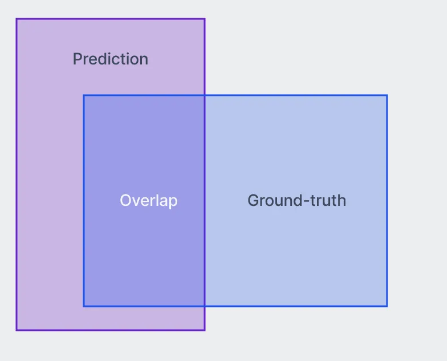}
    \caption{Mathematical Concept of IoU}
    \label{fig:Mathematical Concept of IoU}
\end{figure}
        \subsection{Experimental Results and Analysis}

This section presents the results of experiments conducted on the PASCAL VOC 2012 dataset\cite{vicente2014reconstructing}, including IoU value statistics for each category and visual analysis of random samples. The designed network structure's effectiveness and advancement are comprehensively evaluated based on both quantitative metrics and visual performance.

\subsubsection{Dataset Overview and Parameter Settings}

PASCAL VOC 2012, a benchmark dataset, includes 17,125 original images and their corresponding annotations. It is widely used for object detection, image segmentation comparisons, and model performance evaluations. The dataset provides labeled data for supervised learning in visual tasks and is divided into four major categories: person, common animals, transportation vehicles, and indoor furniture. we have chosen to adopt the linked data\cite{li2023investigation} approach to process dataset. This method is particularly advantageous for enhancing the interoperability and accessibility of our dataset, allowing for a more structured and semantic integration of data sources. By applying this methodology, we aim to facilitate a more robust and dynamic analysis, enabling enhanced data discovery and reuse across various research domains.

The parameter settings for experiments using this dataset are detailed in Table 1. Test data was randomly selected, and the network training utilized stochastic gradient descent with momentum as the optimizer, with learning rate and momentum parameters set to 0.1 and 0.9, respectively.
\begin{table}[h]
\centering
\begin{tabular}{lc}
\hline
\textbf{Parameter} & \textbf{Value} \\ \hline
Epoch limit & 210 \\ \hline
Quantity of validation images & 5813 \\
Learning rate & 0.1 \\
Quantity of training images & 5718 \\
Quantity of test images & 1000 \\
Momentum & 0.9 \\

\end{tabular}
\caption{Hyperparameter Settings for PASCAL VOC 2012 Dataset}
\label{tab:Hyperparameter Settings for PASCAL VOC 2012 Dataset}
\end{table}

Experimental Results Comparison and Analysis
Table 2 shows the IoU for each category in the PASCAL VOC 2012 test set, comparing the proposed method with SegNet.
\begin{table}[h]
\centering
\begin{tabular}{lcc}
\hline
\textbf{Category} & \textbf{Proposed Method} & \textbf{SegNet} \\ \hline
Keychain & 92.61 & 89.8 \\
Laptop & 60.23 & 39.2 \\
Window & 93.95 & 79.6 \\
Cup & 74.84 & 63.8 \\
Book & 82.76 & 68.1 \\
Backpack & 95.00 & 87.3 \\
Pen & 88.44 & 81.1 \\
Mouse & 94.61 & 86.0 \\
Desk & 45.42 & 28.4 \\
Jacket & 91.28 & 76.9 \\
Clock & 76.22 & 61.9 \\
Phone & 90.48 & 78.9 \\
Hat & 91.66 & 80.2 \\
Sunglasses & 88.03 & 83.5 \\
Shoe & 87.88 & 80.1 \\
Plant & 69.77 & 58.7 \\
Umbrella & 82.73 & 83.3 \\
Pillow & 60.81 & 54.2 \\
Poster & 80.65 & 80.6 \\
Speaker & 66.73 & 64.9 \\
Mean & 80.71 & 72.4 \\ \hline

\end{tabular}
\caption{Comparison of IOU  for dataset}
\label{tab:Comparison of IOU  for dataset}
\end{table}

In summary, the integration of multiple residual connections enhances the fidelity of the features extracted by the semantic segmentation network, maintaining a higher correlation with the original image. This results in superior pixel-level classification and boundary localization compared to SegNet. Both qualitative visual analysis and quantitative IoU analysis for each category demonstrate that this method effectively leverages the efficiency of max-pooling indices and the flexibility of multiple residual connections. Consequently, it achieves higher accuracy in image semantic segmentation, better meeting practical application requirements.

        \section{Conclusion}  
The challenges posed by the substantial information loss during the multiple down-sampling and up-sampling processes in the SegNet model have been a significant bottleneck in achieving high accuracy in semantic segmentation\cite{sohail2022systematic}. In response, our research introduces a novel encoder-decoder network structure that incorporates multiple residual connections, effectively addressing these limitations. By integrating residual connections, our model harnesses both low-level spatial information and high-level semantic features across various resolutions, thereby preserving essential details that are crucial for accurate segmentation. This strategic use of residual connections not only counters the loss of information but also avoids excessive increases in the number of parameters, maintaining a balance between complexity and performance.

Moreover, recognizing the critical impact of class imbalance on segmentation tasks, we have innovated a balanced cross-entropy loss function. This enhancement optimizes the training process, ensuring more stable and efficient model convergence, and significantly reduces the loss at the point of convergence. Consequently, our approach not only enhances the robustness of the model but also substantially improves segmentation accuracy.

To substantiate our claims, we conducted extensive experimental comparisons and analyses on the PASCAL VOC 2012 dataset. Employing rigorous quantitative evaluation metrics alongside thorough visual analysis, our findings clearly demonstrate that our proposed model markedly outstrips the traditional SegNet in terms of segmentation performance. These results underscore the effectiveness of our structural and functional modifications in advancing the field of semantic segmentation.

\bibliographystyle{IEEEtran}
\bibliography{references}
     
\end{CJK*}	
\end{document}